\documentclass[twocolumn,prb,showpacs,floatfix]{revtex4}

\usepackage{epsfig}
\usepackage{graphics}
\usepackage{graphicx}
\usepackage{dcolumn}
\usepackage{amsmath,amsfonts,amssymb,amsxtra}
\usepackage{mathrsfs}
\usepackage{bm}
\usepackage{times}

\begin{document}

\title{Screening of the Raman response in multiband superconductors -- application to iron-pnictides
}

\author{C. Sauer$^{1}$\footnote{Current address: EP 7, Julius Maximilians Universit\"{a}t W\"{u}rzburg, D-97074 W\"{u}rzburg, e-mail: cesauer@physik.uni-wuerzburg.de \\\\}  and G. Blumberg$^{1}$}

\affiliation{
$^{1}$Department of Physics and Astronomy, Rutgers The State University of New Jersey, Piscataway, NJ 08854 \\\\
}

\thanks{}

\date{\today}

\begin{abstract}
We performed model calculations of Raman responses for multiband 2D superconductors. The multiband effects of screening in the $A_{1g}$ symmetry channel were investigated analytically and numerically for a band structure model mimicing ARPES data on iron-pnictide materials. An acceptable agreement between our model calculations and recent experimental data is demonstrated by modification of the band structure parameters.
\end{abstract}


\maketitle

\section{Introduction}

A large variety of compounds belonging to the iron-based superconductors have been found since the discovery of superconductivity in $LaFeAsO_{1-x}F_x$ in the beginning of 2008 \cite{kamiharadisc}. They can be classified into four different families, divided by stoichiometric differences, exhibiting superconducting (SC) behavior induced by hole-doping, electron-doping as well as solely the application of high pressure (for a review see \cite{reviewishida}). All compounds are layered materials with a common layer of an iron-pnictide in between different spacing layers or simply layers of $FeSe$. So the location of the SC inducing physics is the tetragonal $Fe$-containing layer leading to highly two dimensional (2D) electronic structure \cite{singhcylfsdft1}. Angle resolved photo-emission spectroscopy (ARPES) finds several bands crossing the Fermi surface \cite{arpesnature,arpespnictidenature,arpespnictidegap} (FS) which is a novelty since, with the exception of MgB$_{2}$, superconductors have been most often treated as single band materials. The mechanism that leads to superconductivity including the pairing symmetry of the SC order parameter is still under debate (a theoretical review is provided by Ref. \cite{mazintheoreticalreview}). Gap symmetries both with nodes ($p$-wave or $d$-wave) and without nodes ($s$-wave) are still under consideration \cite{mazintheoreticalreview}.

Electronic Raman scattering (for a review see \cite{reviewraman}) could clarify some of those unresolved issues. In the light scattering experiment different parts of the Brillouin zone (BZ) can be emphasized with different weights through the selection of incident and outgoing light polarizations with respect of the crystallographic directions. Thus by manipulating the directions of the polarization vectors of the incident and scattered photons in a Raman experiment, it is possible to determine the symmetry of the pair-breaking excitations related to the symmetry of the SC order parameter \cite{reviewraman}. Furthermore, the low energy tail of the Raman responses can be used to distinguish between a fully gaped FS or a SC gap with nodes allowing the excitation of quasi particles at arbitrarily small energies.

The theory of electronic Raman spectroscopy in superconductors \cite{kleindierker} has been very successful in applications to high-T$_{c}$ cuprates \cite{reviewraman,strohmcardona} and now allows to make predictions for the Raman responses of the iron-pnictides. Recently model calculations of expected Raman responses for a superconductor with multi-band order parameter and including the long-range Coulomb interactions responsible for the multi-band screening effects have been performed by Boyd et. al. \cite{boydramanpnictide}. These calculations used multiple circular FS sheets as a band structure and for the majority of the work the Raman vertices are obtained by an expansion into FS harmonics for cylindrical FS. Taking the screening of the charged background into account but leaving out vertex corrections capturing effects of final state interactions a variety of different gap distributions on the FS sheets with and without nodes were used. Thus many different forms were obtained for the Raman responses which show the possible scenarios expected from the data. By comparing the singularities in different symmetry channels of such calculated Raman responses with the peak structure of actual Raman data the reconstruction of the momentum and frequency dependence of the superconducting order parameter, critical to understanding of the superconducting mechanism in these materials, is possible. In another study \cite{chubukovgapsym,kleinviewpoint} it has been established by analyses of the vertex corrections that for an $A_{1g}$ extended $s$-wave gap symmetry, the $A_{1g}$ Raman response has a true collective mode resonance peak below the fundamental gap $2\Delta_0$. Since the collective mode is predicted only for this particular pairing symmetry, conscientious polarization resolved low-frequency Raman scattering study of the iron-pnictide materials may provide a way to unambiguously distinguish between various suggested gap symmetries.

Very recently the first electronic Raman scattering experiments on $Ba(Fe_{1-x}Co_x)_2As_2$, a member of the 122-family, have been 
performed \cite{hackldata} showing a change of the Raman response in the SC state only in the $A_{1g}$ and $B_{2g}$ channel. A clear peak of $log$-singularity shape in the latter one and a broader peak in the former one at a slightly higher energy are observed and the low energy shape is interpreted as evidence for accidental gap nodes.

In this paper model calculations of Raman responses for a multiband free electron model are performed. In chapter II the general theory of Raman scattering in the SC state is presented and explicit forms of the Raman vertices and an analytical expression for the Raman susceptibility are calculated for a 2D tetragonal symmetry and a constant SC gap. Then the screening correction is taken into account adding an extra term in the $A_{1g}$ channel. Multiband effects in this symmetry channel for two and more than two bands with an example of a four band response are presented in chapter III. The general effects found in chapter III are used in chapter IV to explain the Raman responses for a band structure model inspired by ARPES data. First a constant gap and then a $k$-dependent gap of extended $s$-wave symmetry is used for which the single band structure is investigated with an only angular dependent model gap. Finally the Raman response with the extended s-wave gap is compared to experimental data and it is shown that the band structure model can be modified to come to acceptable agreement. It is important to note that final state interactions leading to vertex corrections are neglected in this work. The consequences of including vertex corrections that could give rise to resonance modes or notably modify the Raman response are discussed in Refs. \cite{chubukovgapsym,blumbergmgb2}. 
%While in \cite{chubukovgapsym} this leads to the above stated resonance peak a notable shift of Leggett's collective mode is observed in \cite{blumbergmgb2}. So in both cases the corrections due to final state interactions lead to important consequences. 
However, this work focuses on multiband screening effects and features of vertex corrections are left for future publications.

The fact that in the applied assumptions analytical calculations are possible allows us to find the origin of the obtained general mutliband effects in the screened $A_{1g}$ channel. With that the shape of the numerically calculated Raman responses for the particular band structure with extended s-wave symmetry can be explained and further applications to other multiband SC might follow. Furthermore the nature of the main features of the single band Raman responses with the extended s-wave gap are identified analytically with a simplified model gap. Thus our work adds further details to the previous study by Boyd et. al. \cite{boydramanpnictide}.

\section{Theory of electronic Raman scattering in the superconducting state}

The Raman process is inelastic scattering of light in which the energy shift $\omega = \omega_I - \omega_S$ between the incident photon ($\omega_I$, $\textbf{k}_I$) and the scattered photon ($\omega_S$, $\textbf{k}_S$) is measured. In the case of electronic Raman scattering this two photon process creates charge density fluctuations in the sample and trough a virtual interband transition (in the non resonant case) an electron is excited from a state below the Fermi energy ($E_F$) to a state above. In a clean system the energy difference between those two states corresponds to the energy loss of the photon. It can be shown that the cross section of the scattered light is proportional to the imaginary part of the Raman susceptibility \cite{strohmcardona} which is defined in linear response theory in the following way
\begin{equation}
	\chi_{Raman}(\textbf{q}, t) = \frac{i}{\hbar} Tr\left\{ \frac{e^{- \beta H_0}}{Z} \left[ \tilde{\rho}_\textbf{q}(t), \tilde{\rho}_{-\textbf{q}}(0) \right] \right\}
	\label{eq:deframansuscep}
\end{equation}
Here $\textbf{q} = \textbf{k}_I - \textbf{k}_S$, $\beta = k_B T$ ($k_B$ is Boltzmann's constant), $Z$ is the partition function and $H_0$ is the Hamiltonian of the unperturbated system. The perturbation Hamiltonian $H'$ is proportional to the effective density operator \cite{strohmcardona}
\begin{equation}
	\tilde{\rho}_\textbf{q} = \sum_{n, \textbf{k}} \gamma_n(\textbf{k}) c_{n, \textbf{k}}^{\dagger} c_{n, \textbf{k}}^{\:}
	\label{eq:effdensoper}
\end{equation}
which contains fermionic annihilation $(c_{n, \textbf{k}}^{\:})$ and creation $(c_{n, \textbf{k}}^{\dagger})$ operators in the $n^{th}$ band. Furthermore $\tilde{\rho}_\textbf{q}$ is proportional to the so called Raman vertex $\gamma_n(\textbf{k})$ which depends on the polarization direction of the incident and scattered photon and thus on the experimental setup (light polarizers). This quantity will be specified and discussed in detail in the next paragraph. With the wavelength of the incident and scattered photon being in or near the visible spectrum the corresponding wave vectors and therefore also their difference $\textbf{q}$ is much smaller than $\frac{2 \Delta}{\hbar \: v_{F,\textbf{q}}}$ ($2 \Delta$ is the SC gap and $v_{F,\textbf{q}}$ is the component of the Fermi velocity along q). This makes the $\textbf{q} \rightarrow 0$ limit a reasonable simplification that leads to the explicit form of the Raman susceptibility in the SC state
\begin{equation}
	\chi_{\gamma \gamma, n}(\textbf{q} \rightarrow 0, \omega) = \frac{1}{N} \sum_\textbf{k} \gamma^2_n(\textbf{k}) \lambda_n(\textbf{k},\omega).
	\label{eq:absusceptibil}
\end{equation}
The quantity $\lambda_n(\textbf{k},\omega)$ is the Tsuneto function \cite{tsuneto}
\begin{equation}
\begin{aligned}
	\lambda_n(\textbf{k},\omega) = \frac{\Delta^2_n(\textbf{k})}{E^2_n(\textbf{k})} & \tanh\left( \frac{E_n(\textbf{k})}{2 k_B T} \right) \left( \frac{1}{2 E_n(\textbf{k}) + \hbar \omega + i \alpha} \right. \\ &+ \left. \frac{1}{2 E_n(\textbf{k}) - \hbar \omega - i \alpha} \right)
	\label{eq:tsunetofuncgen}
\end{aligned}
\end{equation}
with the SC gap $\Delta_n(\textbf{k})$, the SC quasi particle energy $E_n(\textbf{k}) = \sqrt{\epsilon^2_n(\textbf{k}) + \Delta^2_n(\textbf{k})}$ and a small broadening factor $\alpha$. This function represents the creation of quasi particles out of the SC condensate. The first term is proportional to a SC coherence factor and is unity at the Fermi momentum. The further away from $E_F$ the quasi particles are created the smaller this factor gets so the major contribution to the Raman response originates from the vicinity of the $E_F$. The second term is a statistical factor that takes into account already thermally excited quasi particles. Since the following calculations are made under the assumption that $T=0$ this factor is simply unity. The third term takes care of the necessary energy conservation and excludes a creation of quasi particles inside the SC gap.

Now the Raman vertex will be specified by an analytical expression. In the nonresonant case, where the electronic excitation does not include a real intermediate state, the Raman vertex can be calculated with an effective mass approximation \cite{strohmcardona} in the following form
\begin{equation}
	\gamma_n(\textbf{k}) = \frac{m}{\hbar^2} \sum_{i, j} \textbf{e}_{S, i}^* \frac{\partial^2 \epsilon_{n}(\textbf{k})}{\partial k_i \partial k_j} \textbf{e}_{I, j}^{\:}
	\label{eq:effmassramvertex}
\end{equation}
with $m$ denoting the electrons mass and $\textbf{e}_{S, i}^*$ and $\textbf{e}_{I, j}^{\:}$ being the light polarization vectors of the scattered and incident photon, respectively. In this approximation the Raman vertex is proportional to the second derivative (curvature) of the band energy dispersion in the particular $k$-direction selected by the two light polarization vectors which is proportional to the inverse effective mass of the band in $\textbf{k} \cdot \textbf{p}$ - theory. Taking into account that a periodic crystal transforming according to a special point symmetry group is investigated group theoretical arguments can be used to further simplify the expression in Eq. (\ref{eq:effmassramvertex}). Any operator of the form $M_{f,i}=M_{i,f}^{\alpha,\beta} e_I^\alpha e_S^\beta$ ($i$=initial, $f$=final) can be decomposed according to the point symmetry group of the crystal which will be taken to be $D_{4h}$ in this case. This leads to the decomposition of $M_{f,i}$ into projected operators $O_{\nu}$ corresponding to the irreducible representation $\nu$ of the point group \cite{reviewraman}.
\begin{equation}
\begin{aligned}
	M_{f,i} &= \frac{1}{2} O_{A_{1g}} (e_I^x e_S^x + e_I^y e_S^y) + \frac{1}{2} O_{B_{1g}} (e_I^x e_S^x - e_I^y e_S^y)\\  &+ \frac{1}{2} O_{B_{2g}} (e_I^x e_S^y + e_I^y e_S^x) + \frac{1}{2} O_{A_{2g}} (e_I^x e_S^y - e_I^y e_S^x)
\end{aligned}
\end{equation}
which results, together with Eq. (\ref{eq:effmassramvertex}), in the Raman vertices for the four symmetry channels
\begin{equation}
\begin{aligned}
	\gamma_{n,A_{1g}} =& \frac{m}{2 \hbar^2} \left( \frac{\partial^2 \epsilon_{n}(\textbf{k})}{\partial k_x \partial k_x} + \frac{\partial^2 \epsilon_{n}(\textbf{k})}{\partial k_y \partial k_y} \right)\\
	\gamma_{n,B_{1g}} =& \frac{m}{2 \hbar^2} \left( \frac{\partial^2 \epsilon_{n}(\textbf{k})}{\partial k_x \partial k_x} - \frac{\partial^2 \epsilon_{n}(\textbf{k})}{\partial k_y \partial k_y} \right)\\
	\gamma_{n,B_{2g}} =& \frac{m}{2 \hbar^2} \left( \frac{\partial^2 \epsilon_{n}(\textbf{k})}{\partial k_x \partial k_y} + \frac{\partial^2 \epsilon_{n}(\textbf{k})}{\partial k_y \partial k_x} \right)\\
	\gamma_{n,A_{2g}} =& \frac{m}{2 \hbar^2} \left( \frac{\partial^2 \epsilon_{n}(\textbf{k})}{\partial k_x \partial k_y} - \frac{\partial^2 \epsilon_{n}(\textbf{k})}{\partial k_y \partial k_x} \right).\\
	\label{eq:parabolverteces}
\end{aligned}
\end{equation}
The different symmetry channels can be investigated by choosing certain combinations of the light polarization vectors through the setup of light polarizers in the experiment. Using an effective mass approximation (Eq. (\ref{eq:effmassramvertex})) to obtain the Raman vertices of Eq. (\ref{eq:parabolverteces}) is a different approach than the one in Boyd et al. \cite{boydramanpnictide} where an expansion into FS harmonics is done leading to angular dependent vertices. In this work a 2D free electron model for the band structure will be used which results in a constant vertex for the $A_{1g}$ channel since the second derivative of the energy band dispersion is proportional to the constant inverse effective mass $\frac{1}{m^*}$ of the band. For a parabolic band all other symmetry channels are vanishing which is a consequence of the effective mass approximation in combination with a free electron model. This leads to the focus of the following investigations being on the fully symmetric $A_{1g}$ channel. A tight binding model, which is the next possible step, will not lead to vanishing vertices for the non fully symmetric channels (except $A_{2g}$, the symmetry of the antisymmetric tensor \cite{strohmcardona}, which is usually negligible).

Using a 2D free electron model for the energy dispersion of the $n^{th}$ band, which is determined by the effective mass $m_n^*$ and the chemical potential $\mu_n$, the Raman susceptibility $\chi_{\gamma \gamma, n}$ can be calculated analytically. First the Tsuneto function of Eq. (\ref{eq:tsunetofuncgen}) is inserted into Eq. (\ref{eq:absusceptibil}), the limit of $\alpha \rightarrow 0$ is taken and the sum is replaced by a 2D $k$-integration. Evaluating only the imaginary part of the integral and inserting the result into the Kramers-Kr\"{o}nig relation leads to the analytical expression
\begin{equation}
	\chi_{\gamma \gamma,n}(\omega) = N_{F,n} \left(\gamma_n\right)^2 \frac{\arcsin(x_n)}{x_n \sqrt{1 - x_n^2}} = N_{F,n} \left(\gamma_n\right)^2 F_n
	\label{eq:analyticalform}
\end{equation}
with the definition
\begin{equation}
	F_n = \frac{\arcsin(x_n)}{x_n \sqrt{1 - x_n^2}}
	\label{eq:definitionfn}
\end{equation}
and with $x_n=\frac{\omega}{2 \Delta_n}$ and $N_{F,n}$ being the 2D density of states (DOS) at $E_F$ which is a constant proportional to $\left|m_n^*\right|$ (detailed calculation can be found in \cite{masterthesis}). Since the cross section of the Raman process or Raman efficiency is proportional to $Im[\chi_{\gamma \gamma, n}]$ the one band response is obtained by taking the imaginary part of Eq. (\ref{eq:analyticalform}). For $0 \leq x_n < 1$ ($0 \leq \omega < 2 \Delta_n$) $Im[\arcsin(x_n)]$ is vanishing and the square root is entirely real leading to a vanishing Raman response. This is due to the fact that no quasi particles can be created without breaking a Cooper pair which has a binding energy of $2 \Delta_n$. At $\omega = 2 \Delta_n$ ($x_n=1$) the denominator hits zero and the function diverges as a square root singularity, a consequence of the diverging SC quasi particle DOS at this energy. Above $\omega = 2 \Delta_n$ ($x_n > 1$) the square root is entirely imaginary and in $Im[\chi_{\gamma \gamma, n}]$ only $Re[\arcsin(x_n)]$ remains which is a constant of $\frac{\pi}{2}$ leading to $Im[\chi_{\gamma \gamma, n}] = \frac{\pi}{2} Re[\frac{1}{x_n \sqrt{x_n^2 -1}}]$ for the single band Raman response.

So far it was not taken into account that in the solid a charged background of electrons is present that reacts through long range Coulomb interaction on the charge density fluctuations imposed by the Raman process. The electron background rearranges in a way to screen the perturbative electric fields and thus works against the Raman process. This screening effect can be included in the theory within a random phase approximation like sum \cite{strohmcardona} that leads in the limit of small $\textbf{q}$ (compared to the inverse coherence length $\xi$ and the Fermi wave vector $k_F$) to an additional term in the Raman susceptibility. This term includes a $\textbf{k}$ sum over the 1st BZ of $\gamma(\textbf{k}) \cdot \lambda(\textbf{k})$ that vanishes if $\gamma(\textbf{k})$ is not fully symmetric hence exhibiting sign changes in between different parts of the 1st BZ leading to those parts canceling out each other. Thus only the fully symmetric $A_{1g}$ channel gets screened and the screened Raman susceptibility is
\begin{equation}
	\chi_{Raman}(\omega) = \sum_n \chi_{\gamma \gamma,n}(\omega) - \frac{\left(\sum_n \chi_{\gamma 1,n}(\omega)\right)^2}{\sum_n \chi_{1 1,n}(\omega)}
	\label{eq:includescreen}
\end{equation}
with
\begin{equation}
	\chi_{\gamma 1, n}(\omega) = N_{F,n} \gamma_n \frac{\arcsin(x_n)}{x_n \sqrt{1 - x_n^2}}
	\label{eq:absusceptibil2}
\end{equation}
and
\begin{equation}
	\chi_{1 1, n}(\omega) = N_{F,n} \frac{\arcsin(x_n)}{x_n \sqrt{1 - x_n^2}}.
	\label{eq:absusceptibil3}
\end{equation}
The second term in Eq. (\ref{eq:includescreen}) is called screening term and consists of a normalization by a sum over terms shown in Eq. (\ref{eq:absusceptibil3}) and the square of a sum over terms shown in Eq. (\ref{eq:absusceptibil2}). Considering only one free electron band the Raman response vanishes.
\begin{equation}
\begin{aligned}
	Im\left[\chi_{Raman}(\omega)\right] &= Im\left[ N_{F,n} \gamma^2_n F_n\right] - Im\left[ \frac{\left( N_{F,n} \gamma_n F_n \right)^2}{N_{F,n} F_n }  \right]\\ &= 0
	\label{eq:genformoneband}
\end{aligned}
\end{equation}
This means that the screening of a free electron background is perfect and all charge density fluctuations are completely compensated by the rearrangement of the free electrons.

\section{Multiband effects in the screened $A_{1g}$ channel with constant SC gaps}

\subsection{Two bands in $A_{1g}$ symmetry}

In contrast to the rather trivial one band case interesting features appear as soon as two free electron bands are considered. Taking a closer look at the second term in Eq. (\ref{eq:includescreen}) reveals that through the square in the numerator cross terms of the complex functions $\chi_{\gamma 1,n}$ containing real and imaginary parts corresponding to distinct bands contribute to the two band Raman response. Having two bands furthermore adds complexity by the possibility of choosing different effective masses (magnitude and sign) and SC gaps. The general form of the screened $A_{1g}$ Raman response shown in Eq. (\ref{eq:genformoneband}) extended for two bands can be transformed into a simpler form
\begin{equation}
\begin{aligned}
	Im&\left[\chi_{Raman}(\omega)\right] = Im\left[ N_{F,1} \gamma^2_1 F_1 + N_{F,2} \gamma^2_2 F_2 \right]\\ &- Im\left[ \frac{\left( N_{F,1} \gamma_1 F_1 + N_{F,2} \gamma_2 F_2 \right)^2}{N_{F,1} F_1 + N_{F,2} F_2} \right]\\ &= Im\left[ \left( \gamma_1 - \gamma_2 \right)^2 \frac{N_{F,1} F_1 N_{F,2} F_2}{N_{F,1} F_1 + N_{F,2} F_2} \right]
	\label{eq:genramrespwithscree}
\end{aligned}	
\end{equation}
For the sake of generality the effective masses (determining $\gamma_n$ and $N_{F,n}$) and the SC gap (present in $F_n$) are taken to be different for both bands performing the transformation in Eq. (\ref{eq:genramrespwithscree}). We note here that for a two band model the vertex corrections due to final state interactions can notably modify the Raman response (compare Eq. (\ref{eq:genramrespwithscree}) to Eq. (6) in Ref. \cite{blumbergmgb2}). The first thing that becomes obvious in the form of the last line of Eq. (\ref{eq:genramrespwithscree}) is that for equal effective masses (equal $\gamma_n$'s) the two band Raman response is vanishing independent of the choice for the two SC gaps. So the screening is again perfect.

\begin{figure}[ht]
	\centering
		\includegraphics[width=8.5cm]{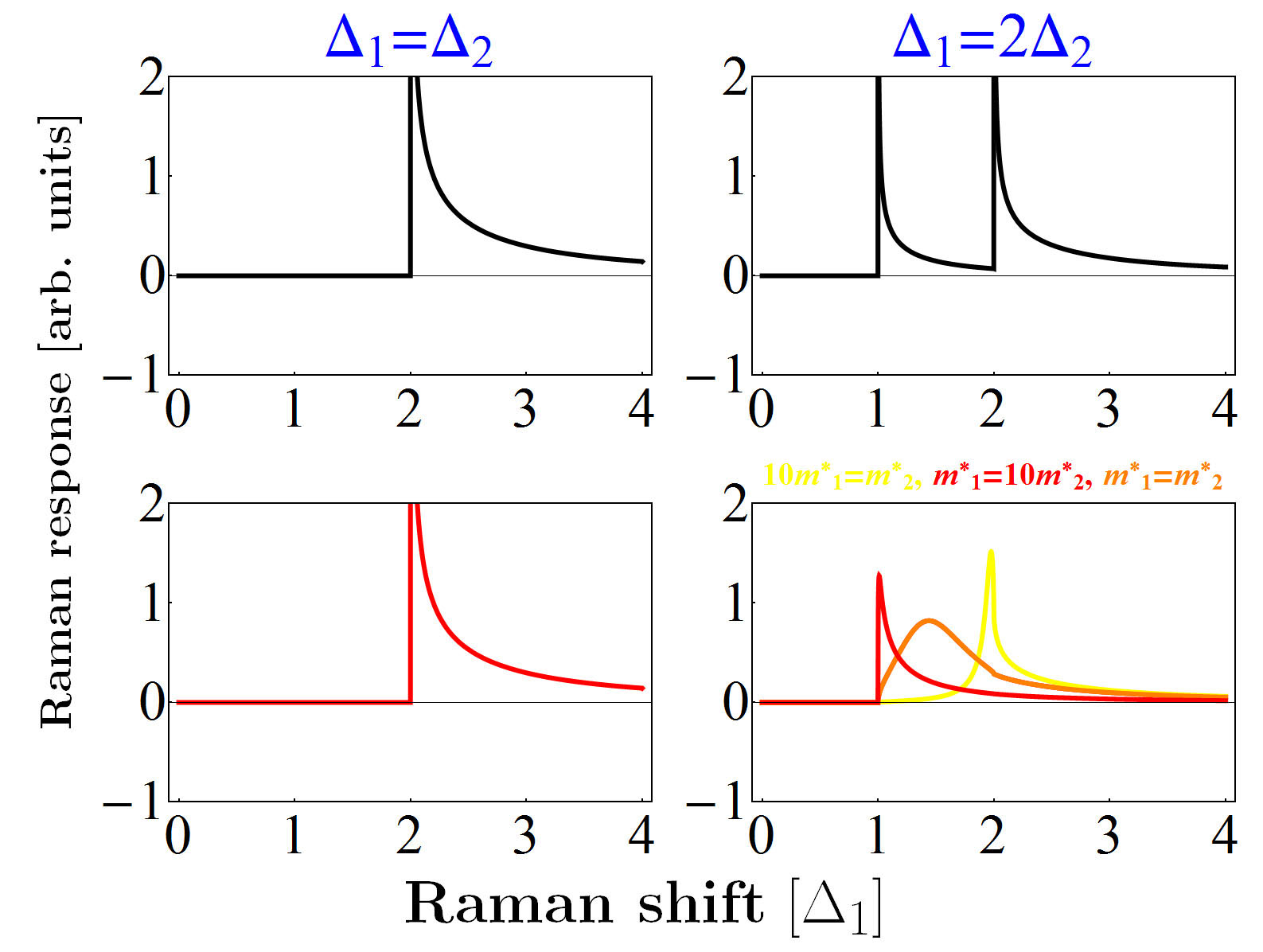}
	\caption{Comparison of unscreened (black) and screened (red) Raman responses in $A_{1g}$ symmetry for two free electron bands with unequal Raman vertices. In the left column the case of two equal constant SC gaps $\Delta_n$ and opposite signs of the effective masses is shown for which the screening is vanishing. In the right column the case of two different SC gaps is displayed with different combinations of effective masses $m^*_1$ and $m^*_2$ for the two bands. In all three cases the singularities at twice the SC gap values are removed. Color online.}
	\label{fig:ConstantGap2Bands}
\end{figure}
Further investigations of the different two band features require a case by case study of all possible combinations of the two tunable quantities, namely the effective mass and the SC gap. First the case of two equal SC gaps ($F_1 = F_2 = F$) is considered. In order to see the effect of screening in this case one has to take a closer look at the screening term which is 
\begin{equation}
\begin{aligned}
	Im\left[ \frac{\left( N_{F,1} \gamma_1+ N_{F,2} \gamma_2\right)^2}{N_{F,1}+ N_{F,2}} F\right] = C \frac{Im[F]}{M^*}
	\label{eq:screningterm}
\end{aligned}	
\end{equation}
with
\begin{equation}
	\frac{1}{M^*} = \frac{ \left( \left|m^*_1\right| \frac{1}{m^*_1} + \left|m^*_2\right| \frac{1}{m^*_2} \right)^2 }{\left|m^*_1\right| + \left|m^*_2\right|}.
	\label{eq:effscreenmass}
\end{equation}
In Eq. (\ref{eq:screningterm}) all common constants are absorbed in $C$. The absolute values of the effective masses in Eq. (\ref{eq:effscreenmass}) have their origin in the fact that the DOS is always a positive quantity in contrast to the Raman vertices for which the sign of the effective mass remains. In the numerator of the inverse effective screening mass $\frac{1}{M^*}$ the absolute value of the effective masses cancels out leaving only a sign dependence. This feature has its origin in the two dimensionality of the band structure for which the DOS is a constant proportional to $|m^*|$ and will not remain for 1D or 3D systems where the DOS includes an energy dependence. For two bands with equal signs this leads to a finite screening term and thus a partial screening (if the effective masses have different absolute values) while for unequal signs the numerator of $\frac{1}{M^*}$ in Eq. (\ref{eq:effscreenmass}) is zero. Thus for a combination of a hole-like and an electron-like band with equal SC gaps the screening term vanishes and the two band Raman response is entirely unscreened no matter how different the absolute values of the effective masses of the two bands are. This case is displayed in the left column of Fig. \ref{fig:ConstantGap2Bands} where the unscreened Raman response (black) and the screened one (red) show no difference demonstrating the vanishing screening term. Additionally it has to be noted that the square root singularity previously found in the unscreened one band Raman response remains in the screened two band Raman response for equal SC gaps.

Now the case of unequal SC gaps is investigated. Reinserting the explicit form of $F_n$ (see Eq. (\ref{eq:definitionfn})) into the simplified form of the screened two band Raman response (last line in Eq. (\ref{eq:genramrespwithscree})) results in 
\begin{equation}
	Im\left[\left(\gamma_1 - \gamma_2 \right)^2 \frac{N_{F,1} N_{F,2} AS_1 AS_2}{N_{F,1} AS_1 SQ_2 + N_{F,2} AS_2 SQ_1} \right]
	\label{eq:insertedbackgenform}
\end{equation}
with the new short representations
\begin{equation}
\begin{aligned}
	AS_n &= \arcsin\left( x_n \right)\\ SQ_n &= x_n\sqrt{1 - x_n^2}.
\end{aligned}	
\end{equation}
The form in Eq. (\ref{eq:insertedbackgenform}) shows that the singularities at $\omega = 2\Delta_{1,2}$, present in the unscreened case, are both removed in the screened case since the square root terms $SQ_1$ and $SQ_2$ hit zero separately always leaving one of them finite. This case is illustrated in the right column of Fig. \ref{fig:ConstantGap2Bands} with different combinations of effective masses for the screened Raman response (red, orange, yellow). For equal effective masses (orange) a broad dome, peaked in the middle of twice the two SC gap values, appears. When both bands have different effective masses (red, yellow) this shape changes into a sharper peak located in the vicinity of twice the SC gap corresponding to the band with the smaller effective mass.

\subsection{More than two bands in $A_{1g}$ symmetry}

Having seen the interesting features originating from the interplay of two bands through their cross terms in the screening term the next logical step is to investigate what happens by including more bands into the Raman response. Writing down a general form for more than two bands (analog to the upper two lines of Eq. (\ref{eq:genramrespwithscree})) it becomes clear that the screening vanishes only in very special cases. Only if all SC gaps are equal and an even number of bands consisting of the same amount of hole-like and electron-like bands is considered the Raman response will be unscreened independent of the absolute values of the effective masses. In all other cases the strength of the screening will depend on the combination of effective masses of the contributing bands. A transformation analog to the one done in Eq. (\ref{eq:genramrespwithscree}) including $n$ free electron bands leads to
\begin{equation}
	Im\left[\chi_{Raman}(\omega)\right] = Im\left[ \sum_{i,j} \left( \gamma_i - \gamma_j \right)^2 \frac{N_{F,i} F_i N_{F,j} F_j}{\left(\sum_l N_{F,l} F_l\right)} \right].
	\label{eq:genformnbands}
\end{equation}
Here $i$,$j$ and $l$ run from $1$ to $n$. This simplified form shows that the screened $n$ band Raman response consists of a sum over all possible combinations of two band terms (with $i \neq j$) each normalized by all $n$ bands. Furthermore it is obvious that terms including two bands with equal Raman vertices vanish and that the dominating terms in the sum are the ones including two bands with opposite signs since for them the Raman vertices in the bracket add up.

\begin{figure}[b]
	\centering
		\includegraphics[width=8.5cm]{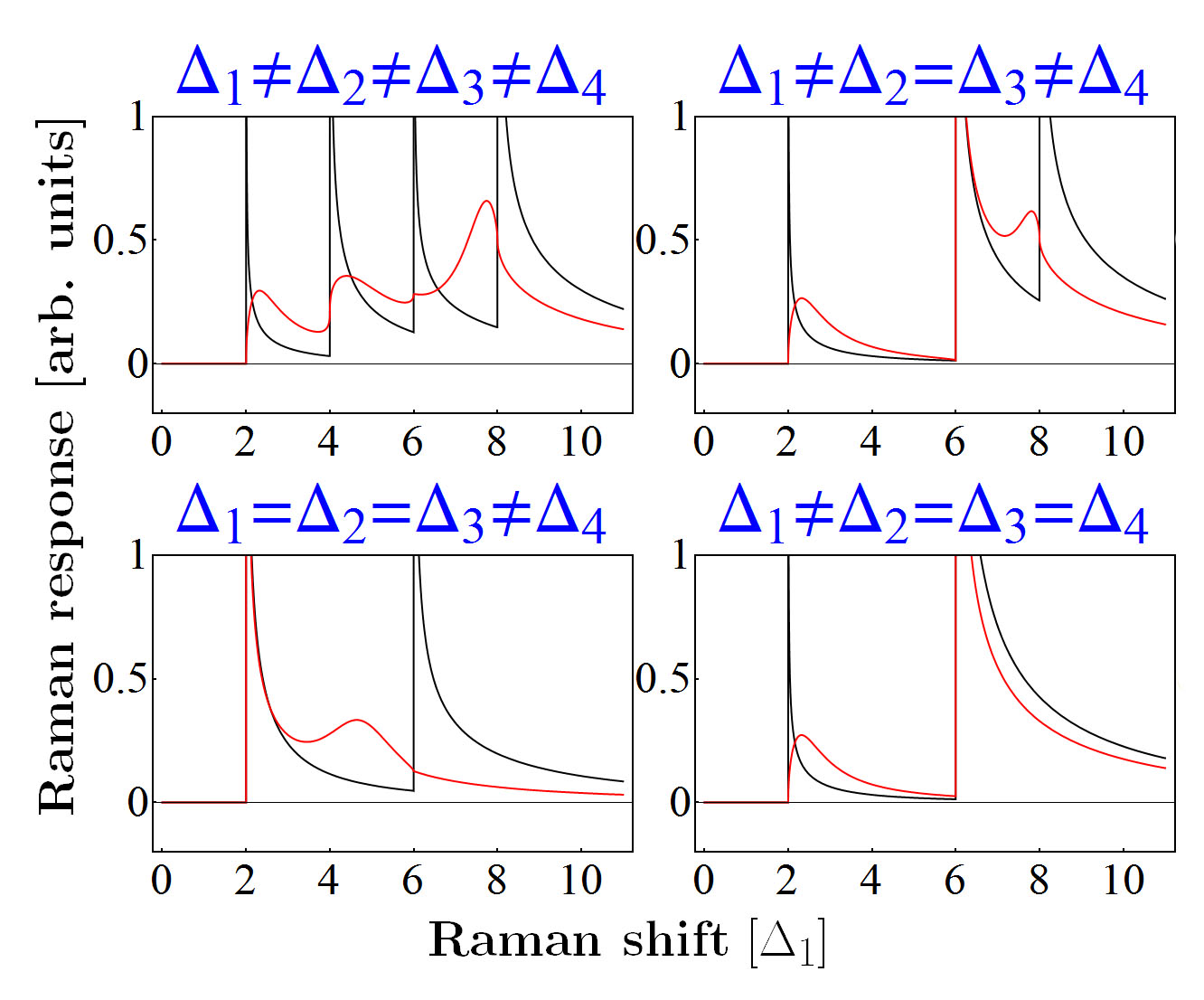}
	\caption{Unscreened (black) and screened (red) Raman response of four free electron bands with constant SC gaps $\Delta_n$ in $A_{1g}$ symmetry. Singularities corresponding to uniquely present gap values are removed by screening and the dome-like shape previously seen in the two band case appears. Color online.}
	\label{fig:ConstantGap4Bands}
\end{figure}
Now the issue of the removal of the singularities previously seen for the two band response with unequal SC gaps will be studied for more than two bands. In order to simplify this investigation the example of a four band Raman response, instead of the general $n$ band form in Eq. (\ref{eq:genformnbands}), will be considered first. For four bands there are six possible combinations of two band terms that add up in the sum of the total Raman response. Each of those six terms can be written as
\begin{equation}
	Im\left[ \left( \gamma_i - \gamma_j \right)^2 \frac{K_i K_j} {K_1 L_1 + K_2 L_2 + K_3 L_3 + K_4 L_4} \right]
	\label{eq:newexplicitfourbandresponse}
\end{equation}
with
\begin{equation}
\begin{aligned}
	K_n &= N_{F,n} AS(n)\\ L_n &= \frac{SQ_i SQ_j}{SQ_n}.
\end{aligned}
\end{equation}
If the two bands $i$ and $j$ in Eq. (\ref{eq:newexplicitfourbandresponse}) have different SC gaps $SQ_i$ and $SQ_j$ will hit zero at different energies. Thus the term $L_i = SQ_j$ will remain finite when $L_j = SQ_i = 0$ and vice versa leading to a finite denominator of Eq. (\ref{eq:newexplicitfourbandresponse}) for all energies. For equal SC gaps in Eq. (\ref{eq:newexplicitfourbandresponse}) $SQ_i$ and $SQ_j$ are equal and will vanish at the same energy resulting in a vanishing denominator and hence a singularity at that energy. This means that as soon as there are two bands with equal SC gaps contributing to the four band Raman response there is one term in the sum that is singular at twice this SC gap. An illustration of this is seen in Fig. \ref{fig:ConstantGap4Bands} where different cases for the SC gaps are considered. The unscreened Raman response (black) shows the location of the singularities and the screened one (red) displays which ones get removed through screening. Whenever a gap is present more than once the screened four band Raman response contains a singularity because of the above given explanation. The generalization of the four to the $n$ band case can easily be done by adding further terms to the denominator of Eq. (\ref{eq:newexplicitfourbandresponse}). The argumentation will remain the same as above and the discovered result that only singularities corresponding to uniquely present gap values are removed holds true.

\section{Application to a multiband free electron band structure based on ARPES data for an iron-pnictide superconductor}

\subsection{Band structure model}

Here the general investigations of free electron Raman responses in the SC state will be applied to a particular band structure model inspired by ARPES measurements. In $Ba_{1-x}K_xFeAs$ the band structure was obtained \cite{arpespnictidenature,arpespnictidegap} consisting of two hole like, almost parabolic bands around the $\Gamma$-point and a propeller-like structure around the $M$-point. This propeller has an electron-like parabolic band in its center which is surrounded by four hole-like blades. In order to mimic this structure within a free electron model the two bands around the $\Gamma$-point and the one around the $M$-point are approximated by a 2D parabola which has the form
\begin{equation}
	\epsilon_{i}(\textbf{k}) = \frac{\hbar^2}{2 m_{i}^*} \left( \left(k_x - g\right)^2 + \left(k_y - h\right)^2 \right) - \mu_{i}
	\label{eq:parabolicband}
\end{equation}
with $g,h \in \mathbb{Z}$. The blades are approximated by 2D ellipsoids, having two different effective masses in the $k_x+k_y$- and $k_x-k_y$-directions, represented by
\begin{equation}
\begin{aligned}
	\epsilon_{bl}(\textbf{k}) &= \frac{\hbar^2}{2 m_{x+y}^*} \left\{ \frac{ \left[ \left( k_x - \delta_1 \right) + \left( k_y - \delta_2 \right) \right]}{\sqrt{2}} \right\}^2 \\  &+ \frac{\hbar^2}{2 m_{x-y}^*} \left\{ \frac{ \left[ \left( k_x - \delta_3 \right) - \left( k_y - \delta_4 \right) \right]}{\sqrt{2}} \right\}^2 - \mu_{bl}.
\end{aligned}
	\label{eq:ellipticalband}
\end{equation}
\begin{figure*}[ht]
	\centering
		\includegraphics[width=17cm]{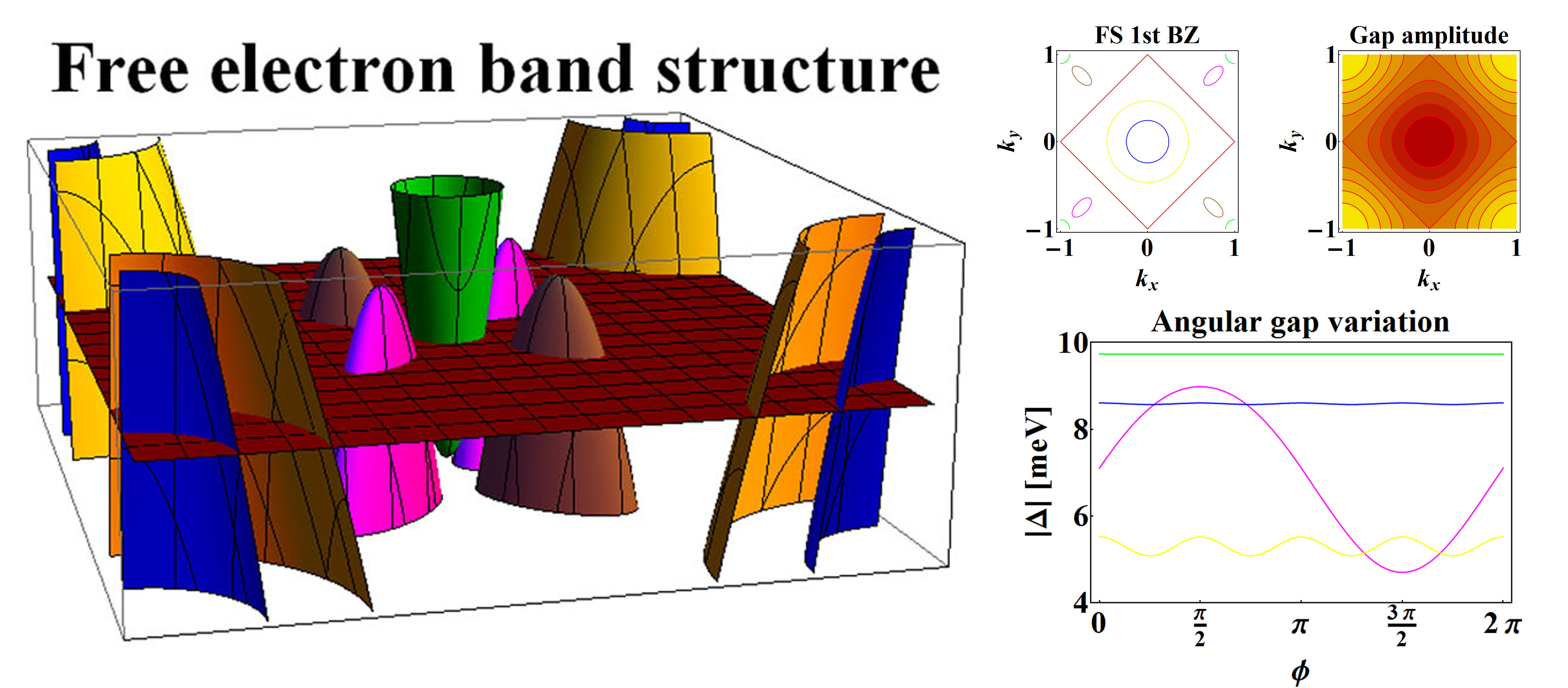}
	\caption{\textbf{Left:} 2D free electron band structure model based on ARPES data \cite{arpespnictidenature,arpespnictidegap}. Four different bands are crossing the FS (red plane). The inner $\Gamma$ parabola (blue) and the outer $\Gamma$ parabola (yellow) are centered around the BZ center ($\Gamma$ point) and a propeller structure consisting of the parabolic $M$-pocket (green) surrounded by four elliptical blades (magenta/brown) is centered around the BZ corner ($M$-point). \textbf{Top middle:} FS topology of the 1st BZ (coloring of bands equal to left panel) together with the node of the extended s-wave gap (red). \textbf{Top right:} Contour plot of the magnitude of the extended s-wave gap given by Eq. (\ref{eq:extendedswave}) with $\Delta_0 = 10 \, meV$. In the middle (BZ center) the gap exhibits its maximum ($+\Delta_0$) and in the corner it reaches its minimum ($-\Delta_0$). \textbf{Bottom right:} Absolute value of the extended s-wave gap at the FS for the different bands (coloring equal to left and top middle panel) as a function of angle $\phi$ (origin of polar coordinates always in the middle of the band). The angular variation of the gap on the inner $\Gamma$ parabola (blue) and $M$-pocket (green) is equal to the one of the outer $\Gamma$ parabola (yellow) but not resolved on this energy scale. The elliptical blade is transformed to a parabolic band for this illustration. Color online.}
	\label{fig:BandStructureAndGap2}
\end{figure*}
Here $\delta_i = \xi_{i,1} + \frac{ 0.346 }{\sqrt{2}}\xi_{i,2}$, with $\xi_{i,1} \in \mathbb{Z}$ and $\xi_{i,2} = -1,0,1$, which is extracted from \cite{arpespnictidenature}. The effective masses and chemical potentials $\mu_n$ used for the different bands are given in TABLE \ref{tab:effmasschempot} \cite{privatcom}. Eq. (\ref{eq:parabolicband}) and Eq. (\ref{eq:ellipticalband}) with the parameters from TABLE \ref{tab:effmasschempot} lead to the band structure shown in the left of Fig. \ref{fig:BandStructureAndGap2} and its corresponding FS displayed in the top middle of Fig. \ref{fig:BandStructureAndGap2}.

\begin{table}[ht]
	\centering
	\caption{Effective masses and chemical potentials used for Fig. \ref{fig:BandStructureAndGap2}.}
		\begin{tabular}{|c||c|c|c|c|c|}\hline
			 {\bfseries band}	& inner $\Gamma$-par. & outer $\Gamma$-par. & $M$-pocket & blade \\\hline
			{\bfseries $m^*$ [$m_e$]} & -2.1 & -6 & 0.8 & -0.61, -2.46 \\\hline
			{\bfseries $\mu$ [$meV$]} & 35 & 45 & -17 & 10 \\\hline
		\end{tabular}
		\label{tab:effmasschempot}
\end{table}

Combining Eq. (\ref{eq:parabolverteces}) with Eq. (\ref{eq:parabolicband}) and Eq. (\ref{eq:ellipticalband}) results in explicit Raman vertices for the used band structure model. For the parabolic bands we obtain
\begin{equation}
\begin{aligned}
	\gamma_{A_{1g}}^i =& \frac{m}{m_{i}^*}\\
	\gamma_{B_{1g}}^i =& \gamma_{B_{2g}}^i = \gamma_{A_{2g}}^i = 0\\
	\label{eq:parabolverteces2}
\end{aligned}
\end{equation}
and the Raman vertices for the elliptical bands are
\begin{equation}
\begin{aligned}
\gamma_{A_{1g}}^{bl} =& \frac{m}{2} \left( \frac{1}{m^*_{x+y}}\right. + \left.\frac{1}{m^*_{x-y}} \right) \\
	\gamma_{B_{2g}}^{bl} =& \frac{m}{2} \left( \frac{1}{m^*_{x+y}}\right. - \left.\frac{1}{m^*_{x-y}} \right) \\
	\gamma_{A_{2g}}^{bl} =& \gamma_{B_{1g}}^{bl}  = 0. \\
	\label{eq:ellipsoverteces}
\end{aligned}
\end{equation}
For the used band structure all bands contribute to the fully symmetric $A_{1g}$ channel while the $B_{2g}$ channel only has a contribution from the elliptical blades and the other symmetry channels completely vanish. This is a consequence of the free electron model in combination with the effective mass approximation used inhere.

At this point it is interesting what the new aspects are that come into play by including elliptical bands. The effect that is a direct consequence of the form of the Raman vertex and the 2D DOS, both different for parabolic and elliptical bands, is the vanishing screening for two bands of opposite signs and equal constant SC gaps. Since the Raman vertex of an elliptical band in $A_{1g}$ symmetry is not just proportional to the inverse effective mass but to the sum of the two different effective masses in the particular $k$-direction and the 2D DOS is proportional to the square root of the product of the two effective masses the inverse effective screening mass differs from Eq. (\ref{eq:effscreenmass}). For two elliptical bands it changes to
\begin{equation}
	\frac{1}{M^*_{ell}} = \frac{ \left( \sqrt{\frac{m^*_{12}}{ m^*_{11}}} \left( 1 + \frac{m^*_{11}}{m^*_{12}} \right) \pm \sqrt{\frac{m^*_{22}}{ m^*_{21}}}  \left( 1 + \frac{m^*_{21}}{m^*_{22}} \right) \right)^2}{\sqrt{m^*_{11} m^*_{12}} + \sqrt{m^*_{21} m^*_{22}}}
	\label{eq:transparenteffmassterm}
\end{equation}
where the plus describes two bands of the same sign and the minus two bands of opposite signs and $m^*_{x+y}$ and $m^*_{x-y}$ for band $n$ are denoted by $m^*_{n1}$ and $m^*_{n2}$ for a shorter notation. Eq. (\ref{eq:transparenteffmassterm}) shows that for two elliptical bands with equal signs the screening is not vanishing (just like it is the case for two parabolic bands) and that its strength depends on the absolute values of the effective masses. Furthermore it can be seen that for two bands of opposite signs and different ratios between $m^*_{n1}$ and $m^*_{n2}$ the screening term remains finite while for the same ratio of the effective masses both terms in the numerator of Eq. (\ref{eq:transparenteffmassterm}) cancel each other and the screening is vanishing.

\subsection{Screened $A_{1g}$ Raman response with a constant gap}

In order to be able to calculate the Raman response of the introduced band structure shown in Fig. \ref{fig:BandStructureAndGap2} the distribution of the SC gap size along the different bands has to be specified. As a first step the SC gaps are assumed to be constant on each band which makes it possible to directly apply the general investigations of chapter III. ARPES measurements find two different SC gaps \cite{arpespnictidegap} of which the smaller one is located on the outer $\Gamma$-parabola and the bigger one is present on all the other bands. The magnitude found for the smaller gap is roughly half the size of the bigger one. Thus the distribution of the SC gaps used here consists of a smaller gap on the outer $\Gamma$-parabola and a bigger one on the inner $\Gamma$-parabola, the $M$-pocket and the blades both related by $2\Delta_1 = \Delta_2$. 
\begin{figure}[ht]
	\centering
		\includegraphics[width=8.5cm]{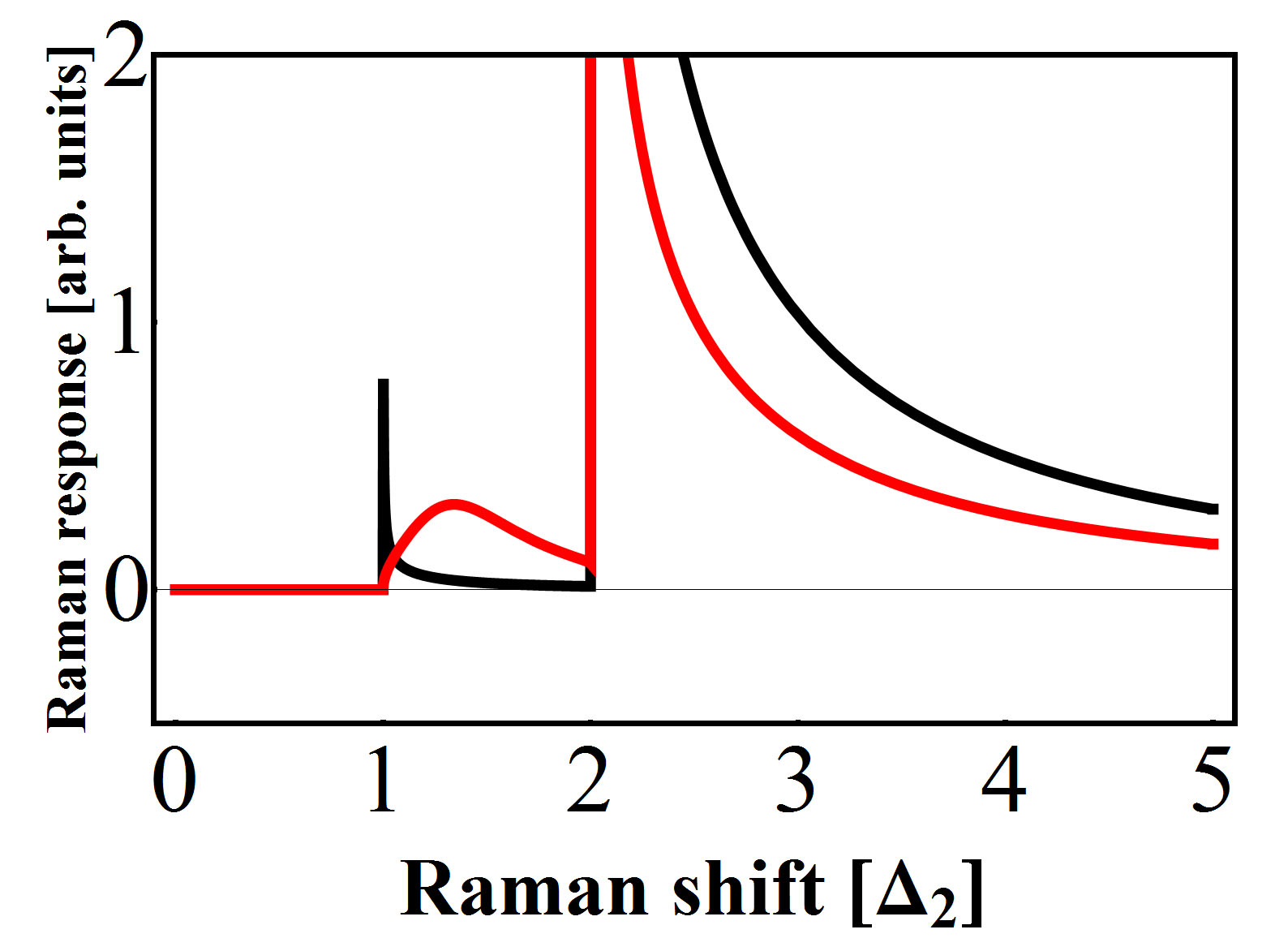}
	\caption{Unscreened (black) and screened (red) Raman response in $A_{1g}$ symmetry of the entire 1st BZ shown in Fig. \ref{fig:BandStructureAndGap2} with two different constant SC gaps. The square root singularity of the smaller gap (present on the outer $\Gamma$-parabola) is removed by screening while the one at the bigger gap (present on all other bands) remains. Color online.}
	\label{fig:ConstantGap1stBZ}
\end{figure}
The screened (red) and unscreened (black) $A_{1g}$ Raman response for these assumptions presented in Fig. \ref{fig:ConstantGap1stBZ} is reminiscent of the one shown in the lower right of Fig. \ref{fig:ConstantGap4Bands}. In both plots the singularity at twice the smaller SC gap is removed since it is uniquely present and the one at twice the bigger SC gap remains because three bands exhibit this same gap value. As previously mentioned the $B_{2g}$ Raman response just has a contribution by the blades leading to a simple square root singularity structure like in the left hand side of Fig. \ref{fig:ConstantGap2Bands} with the singularity located at twice the bigger SC gap. The other two symmetry channels have vanishing Raman vertices and hence show no response at all in this model.

\subsection{$A_{1g}$ Raman response with an extended s-wave gap}

A next step in the calculation of Raman responses for the chosen band structure is to drop the assumption of constant SC gaps and to use one general $k$-dependent SC gap that determines the gap magnitude on each band dependent on its location in the 1st BZ. Experimental results are at this point still controversial even for compounds of the same stoichiometric family (122 in this case) with evidence for gap nodes in some experiments \cite{nodesin122a} and measurements of finite gaps having no nodes in other experiments \cite{nonodesin122,arpespnictidegap}. However the candidate favored by the majority of publications, for example \cite{chubukovgapsym,korshunovspmgap,mazinspmgap}, is the extended s-wave gap which is represented by the following analytical expression \cite{chubukovgapsym}
\begin{equation}
	\Delta(\textbf{k}) = \Delta_0 \left(\cos(\pi k_x) + \cos(\pi k_y)\right).
	\label{eq:extendedswave}
\end{equation}
Here $k$ runs from $-1$ to $1$. This gap exhibits a diamond shaped node in the 1st BZ, shown as a red line in the top middle panel of Fig. \ref{fig:BandStructureAndGap2}, which does not coincide with any of the bands at the FS. Thus all bands are fully gaped but there is a variation of the gap magnitude on either one of them. This angular variation on each band at the FS (after a transformation into polar coordinates) is displayed in the bottom right panel of Fig. \ref{fig:BandStructureAndGap2}. The extended s-wave gap has its maximum $\Delta_0$ at the $\Gamma$-point and its minimum of the same magnitude but negative sign ($-\Delta_0$) is located at the $M$-point. Measured values of the maximum gap \cite{arpespnictidegap,Dingarpesgap,muonspinrotgap} range in between $6 \, meV$ and $12 \, meV$ supporting the choice of $\Delta_0 = 10 \, meV$ for the following calculations.

\begin{figure}[t]
	\centering
		\includegraphics[width=8.5cm]{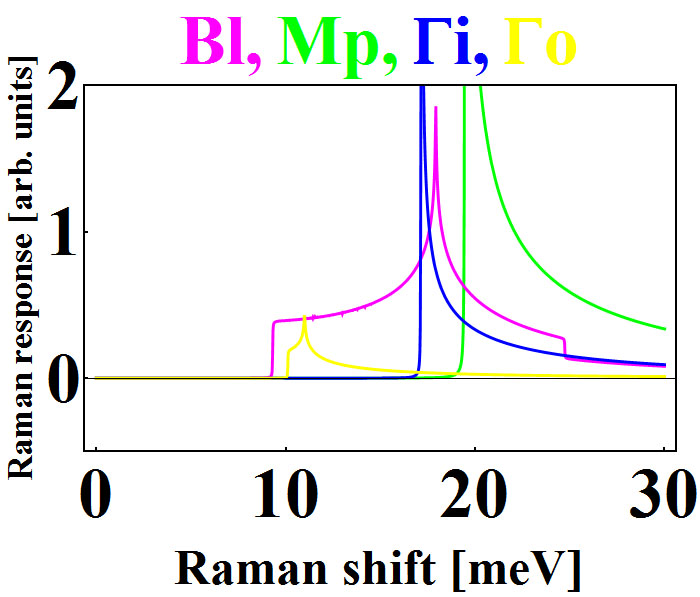}
	\caption{Unscreened single band Raman responses in $A_{1g}$ symmetry of all free electron bands of the band structure displayed in Fig. \ref{fig:BandStructureAndGap2} (same coloring) with extended s-wave gap. All bands show a threshold at twice the minimal gap value and a $\log$-singularity at twice the maximum gap value. These features are only resolved on a smaller energy scale for the inner $\Gamma$ parabola (yellow) and the $M$-pocket (green) because of the smaller variation of the gap magnitude (see Fig. \ref{fig:BandStructureAndGap2}). The blade exhibits an additional discontinuous falloff at higher energies corresponding to the top of the band in the SC state. Color online.}
	\label{fig:SPMSingleBandResponses}
\end{figure}
Combining Eq. (\ref{eq:absusceptibil}) and Eq. (\ref{eq:tsunetofuncgen}) at $T=0$ and noting that the Raman vertex is a constant in the applied model leads to
\begin{equation}
	\chi_{\gamma \gamma,n}(\omega) = \frac{1}{N} \sum_{\textbf{k}}  \gamma_n^2 \frac{4\Delta_n^2(\textbf{k})}{E_n(\textbf{k}) \left[ 4E_n^2(\textbf{k}) - (\omega + i \alpha)^2 \right]}
	\label{eq:imresponsespmgap}
\end{equation}
\begin{figure}[ht]
	\centering
		\includegraphics[width=8.5cm]{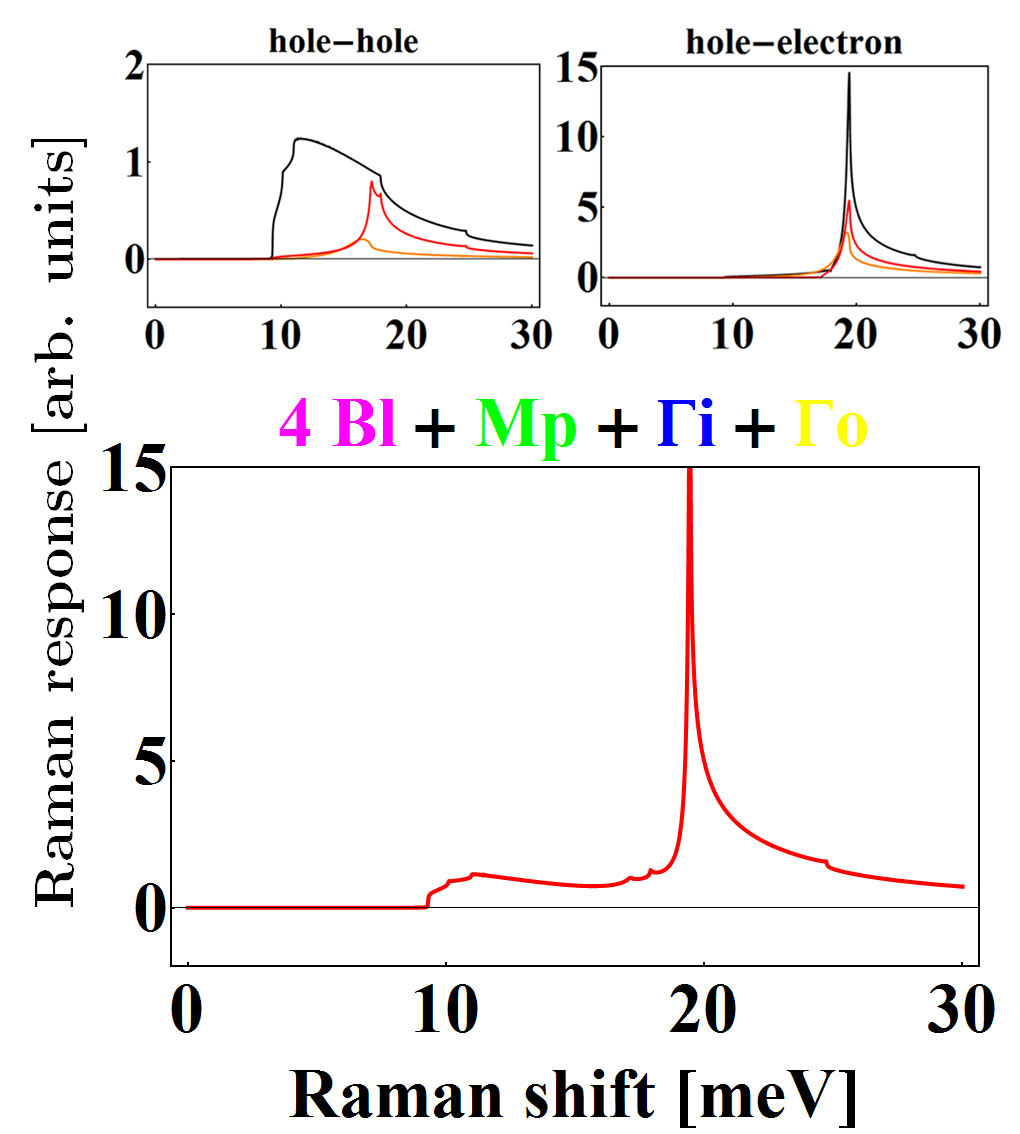}
	\caption{Screened multiband Raman responses in $A_{1g}$ symmetry with extended s-wave gap for combinations of the free electron bands shown in Fig. \ref{fig:BandStructureAndGap2}. \textbf{Top left:} Raman responses for all combinations of two hole-like bands (4 Bl \& $\Gamma$o: black, 4 Bl \& $\Gamma$i: red, $\Gamma$i \& $\Gamma$o: orange). \textbf{Top right:} Raman responses for all combinations of a hole-like and an electron-like band ($Mp$ \& 4 Bl: black, $Mp$ \& $\Gamma$i: red, $Mp$ \& $\Gamma$o: orange). \textbf{Bottom:} Raman response of all four bands of the model. The two band terms shown in the two upper panels renormalized by all four bands sum up to this response resulting in a flat continuum above twice the minimum gap of the blade and a sharp peak near twice the maximum gap of the $M$-pocket. Since all extremal gap values are different for all bands all singularities are removed for the screened two and four band responses. Color online.}
	\label{fig:SPM2and4bandsScreened}
\end{figure}
for the Raman susceptibility of the $n^{th}$ band. For the numerical calculations the sum is replaced by a 2D $k$-integration in polar coordinates and a constant broadening parameter of $\alpha = 0.01 \, meV$ representing a clean system is used. The obtained unscreened single band Raman responses are displayed in Fig. \ref{fig:SPMSingleBandResponses} in the corresponding coloring to the 1st BZ FS in Fig. \ref{fig:BandStructureAndGap2}. The plots show three main features that are seen best in the magenta response of the blade since this is the band exhibiting the largest gap variation at the FS. There is a discontinuous falloff at about $25 \, meV$ corresponding to $\omega = 2\sqrt{\mu_{bl}^2 + \Delta_{bl,center}^2}$ which denotes the top of the band in the SC state above which no quasi particles can be excited. This feature is seen at higher energies for the other bands (since they have larger $\mu$'s) and thus is not present in this plot. At twice the maximum SC gap there is a $log$-singularity and at twice the minimum SC gap a threshold can be seen. These two features are also obtained for the inner $\Gamma$-parabola (yellow) but can only be found at a much smaller energy resolution for the other two bands since those have a much smaller gap variation (see lower right of Fig. \ref{fig:BandStructureAndGap2}). The nature of the $log$-singularity and the threshold are found in analytical investigations of the imaginary part of Eq. (\ref{eq:imresponsespmgap}) using a simplified gap of the form $\Delta(\phi) = \Delta_{off} + \Delta_0 \cos(4 \phi)$ with $\phi$ running from $0$ to $2\pi$. Noting that the integration must only run over the real part of the integrand leads to differences in the integration boundaries at twice the minimum and maximum SC gap resulting in the two distinct features (for a detailed investigation see \cite{masterthesis}).

Knowing the single band Raman response functions is now used to calculate all possible combinations of the screened $A_{1g}$ two band Raman responses and the screened $A_{1g}$ Raman response of all bands. According to Eq. (\ref{eq:genformnbands}) the former ones allow to explain the origin of the features seen in the latter one since it consist of a sum of the renormalized two band terms. In the upper left of Fig. \ref{fig:SPM2and4bandsScreened} the three combinations of two hole-like bands are shown and in the upper right the corresponding plot for the combinations of the three hole-like bands with the electron-like $M$-pocket is illustrated. The total screened $A_{1g}$ Raman response, containing all bands in the 1st BZ, is displayed in the bottom panel of Fig. \ref{fig:SPM2and4bandsScreened}. Its main feature is a sharp peak at about $20 \, meV$ whose origin lies in the three electron-hole two band responses. They all show a peak slightly below twice the maximum SC gap of the $M$-pocket and those add up to the observed peak in the total Raman response. The three hole-hole two band terms generally show a weaker response since the screening is more effective for them leading to a rather featureless broad continuum down to the energy of the threshold of the blade. Since all bands have different crossing lines with the FS the maximum SC gaps are different on all of them and thus all $log$-singularities are removed corresponding to the argumentation in chapter III B. The $B_{2g}$ Raman response again only consists of the four blades which results in a Raman response entirely equal to the magenta single band Raman response of the blade seen in Fig.  \ref{fig:SPMSingleBandResponses} but multiplied by $4$. All other symmetry channels are again vanishing.

\subsection{Comparison with experiment}

The recent measurements of Raman responses \cite{hackldata} on $Ba(Fe_{1-x}Co_x)_2As_2$ only partially resemble the obtained Raman responses with an extended s-wave gap. While the negligible contribution of the SC state to the $B_{1g}$ and $A_{2g}$ response in the experimental data is consistent with the vanishing response in those channels for the used band structure model the $A_{1g}$ and $B_{2g}$ responses are quite different. The structure in the $B_{2g}$ response of the data points towards a power law behavior for small energies and thus supports a SC gap with nodes and not the threshold observed in this work. Even stronger disagreement is seen in the $A_{1g}$ channel where no sharp peak appears in the data but a rather broad dome, peaked at higher energies than the peak in the $B_{2g}$ response, is found. However the $log$-singularity shape of the $B_{2g}$ peak of the here presented Raman response is also present in the data.
\begin{figure}[b]
	\centering
		\includegraphics[width=8.5cm]{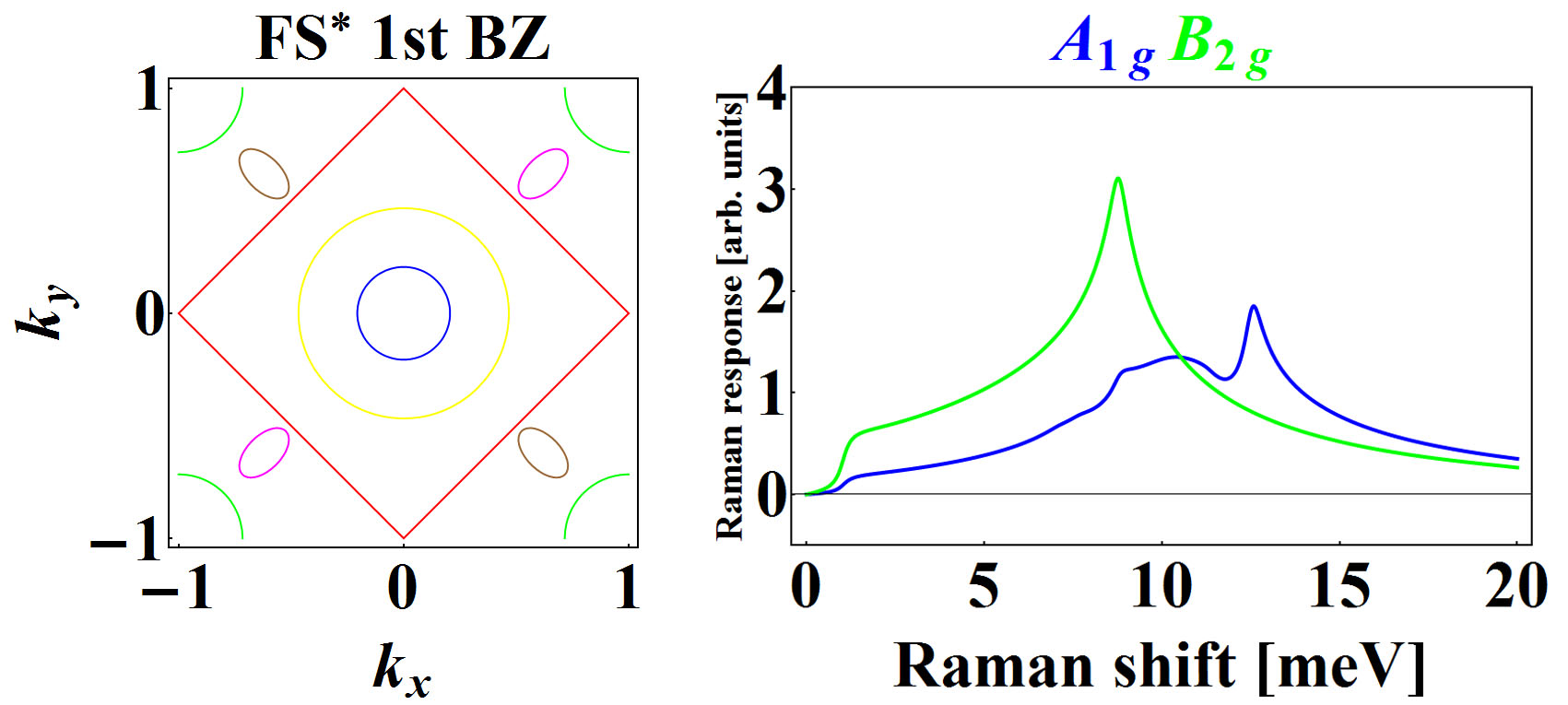}
	\caption{\textbf{Left:} FS$^*$ of 1st BZ of the modified free electron band structure model together with the node of the extended s-wave gap (coloring of bands as in all other figures). \textbf{Right:} Raman responses in $B_{2g}$ (green) and screened $A_{1g}$ (blue) symmetry. The $B_{2g}$ response is multiplied by 5. Color online.}
	\label{fig:NewFSDataMimic}
\end{figure}
One way to tune our band structure model in order to get closer to the measured data is to shift the blades further towards the node line of the gap. To be consistent with the measured peak positions the effective masses of all bands, except the outer $\Gamma$-parabola, and the chemical potential of the $M$-pocket are changed and the maximum gap value is set to $\Delta_0 = 7 \, meV$ . All this and a larger broadening parameter $\alpha = 0.2 \, meV$ leads to the FS topology (FS$^*$) and the Raman responses illustrated in Fig. \ref{fig:NewFSDataMimic}. The threshold in the $B_{2g}$ channel (green) is now at small energies. The total shape is close to the experimental one but to obtain the proper intensity the shown Raman response is multiplied by 5 since the spectral weight of the $B_{2g}$ channel is quite small in the calculated response. This is a consequence of a lack of contribution by other bands than the blades to the $B_{2g}$ channel and might also be due to some resonance effects in the experiment that are not taken into account in the calculations within effective mass approximation. The $A_{1g}$ Raman response (blue) in the right hand side of Fig. \ref{fig:NewFSDataMimic} is now also quite close to the measured one except for the still quite obvious peak at about $13 \, meV$. In general the agreement between the Raman responses of the data on $Ba(Fe_{1-x}Co_x)_2As_2$ and the modified free electron band structure is quite good which shows that the obtained shape of the data does not necessarily have to be explained with gap nodes. Furthermore it is not clear that the measured compound is in the clean limit still leaving the possibility of rather strong impurity scattering that could also explain the low energy tail of the data. With the introduction of possible scattering centers into the $FeAs$ layer (where the important physics for SC is assumed to take place) by the replacement of $Fe$ by $Co$ this scenario is quite reasonable. Certainly more data on other compounds is needed to clarify these issues.

\section{Conclusion}

In this work we performed model calculations of Raman responses for a multiband superconductor in a 2D free electron model. For a constant SC gap parameter analytical calculations were used to investigate multiband effects in the screening correction term in the fully symmetric $A_{1g}$ channel. It was shown that for two parabolic bands with equal SC gaps and opposite signs of effective masses the screening term vanishes which also was proven to hold true for two elliptical bands with equal ellipticity. Moreover the removal of the singularities through screening for two bands with different SC gap sizes was presented. This feature remained in the Raman response of multiple bands for all contributing bands exhibiting a uniquely present SC gap size and the example of a four band response was discussed. Additionally is was shown through analytical calculation that multiband responses consist of a sum of two band terms normalized by all contributing bands.

The general effects found for a constant SC gap parameter were then used to explain the SC Raman response of a model free electron band structure inspired by ARPES measurements on iron-pnictide materials. Using a SC gap of extended s-wave symmetry the single band response features, a threshold at twice the minimum SC gap value and a $\log$-singularity at twice the maximum SC gap value, were clarified. Furthermore we identified that the main features of the four band Raman response of the model band structure with the extended s-wave gap were originating from the two band terms consisting of bands with opposite signs in the effective masses. A comparison of this four band Raman response with experimental data was done and by modifying the band structure parameters an acceptable agreement was demonstrated.

\section*{Acknowledgements}
The authors would like to thank S. Borisenko, D. Evtushinsky and A. Koitzsch for ARPES related discussions and for the band structure parameters.  The work of CS was supported in part by the Deutscher Akademischer Austauschdienst (DAAD).

\end{document}